# Energy efficiency analysis of ammonia-fueled power systems for vehicles considering residual heat recovery

Zexin Nie, Yi Huang, Guangyu Tian

*School of Vehicle and Mobility, Tsinghua University, Beijing, 100084, China*

## Abstract

Ammonia, known as a good hydrogen carrier, shows great potential for use as a zero-carbon fuel for vehicles. However, both the internal combustion engine (ICE) and the proton exchange membrane fuel cell (PEMFC), the currently available engines used by the vehicle, require hydrogen decomposed from ammonia. On-board hydrogen production is an energy-intensive process that significantly reduces system efficiency. Therefore, energy recovery from the system's residual heat is essential to promote system efficiency. ICEs and FCs require different amounts of hydrogen, and they produce residual heat of different quality and quantity, so the system efficiency is not only determined by the engine operating point, but also by the measures and ratios of residual heat recovery. To thoroughly understand the relationships between system energy efficiency and system configuration as well as system parameters, this paper takes three typical power systems with different configurations as our objects. Models of three systems are set up for system energy efficiency analysis, and carry out simulations under different conditions to conduct system output power and energy efficiency. By analyzing the simulation results, the factors that most significantly impact the system efficiency are identified, the guidelines for system design and parameter optimization are proposed.

## 1. Introduction

The extensive reliance on fossil fuels has led to massive greenhouse gas (GHG) emissions. These emissions impose a range of environmental challenges that impact the global ecosystems. Among the major emission sources, the transportation sector constitutes 29% of total greenhouse gas emissions [1]. Therefore, seeking solutions of carbon emission reduction for transportation becomes essential in the global effort to combat climate change.

As a carbon-free alternative fuel, ammonia is recognized as a potential fuel for achieving zero carbon emissions in transportation for the following reasons: ammonia is a good hydrogen carrier[2–5]; ammonia has the third highest volumetric energy density, surpassed only by gasoline and liquefied petroleum gas (LPG)[6,7]; ammonia is the most cost-effective fuel per gigajoule (GJ) when stored onboard[8]; the storage, transportation, and distribution of liquid ammonia are notably accessible, safe and mature[9–12]; ammonia can be used as an internal combustion engine fuel[13–16].

To apply ammonia as a fuel in automotive power systems, on-board hydrogen production from ammonia decomposition is considered a necessary process. Ammonia's slow combustion speed, narrow flammability range and high ignition energy inhibit using it alone in internal combustion engines, so blending ammonia with high-activity fuels is crucial. Hydrogen, with rapid combustion kinetics and broad flammability, stands out as the ideal zero-carbon enhancer[17].There is extensive experimental evidence that ammonia-hydrogen mixtures are well suited for internal combustion engines (ICEs) [18,19]. Furthermore, hydrogen can be obtained on board from ammonia since ammonia is an outstanding hydrogen carrier. Ammonia stores more hydrogen than any other fuel, surpassing even liquid hydrogen [4,20].Therefore, through hydrogen production from ammonia decomposition, both ICEs and proton exchange membrane fuel cells (PEMFCs), which are commonly used in automotive power systems, can be used as engines in ammonia-fueled power systems. There are three typical ammonia-fueled automotive power systems. The first system uses an ammonia-fueled ICE as the engine with an ammonia decomposition unit (ADU) to generate hydrogen [21]. The second system uses a PEMFC as the engine with an ADU and a hydrogen separation unit (HSU) to purify hydrogen [22]. The third system uses both an ammonia-fueled ICE and a PEMFC as engines with an ADU and an HSU [21,23].

For ammonia-fueled power systems applying onboard hydrogen production units, ammonia decomposition reduces systems energy efficiencies. Ammonia decomposition reaction is an endothermic reaction[24,25]. The enthalpy of reaction is 46.1 kJ/mol at standard conditions, which means that 46.1 kJ of thermal energy is consumed when 1mol of ammonia is completely decomposed into hydrogen and nitrogen. However, ammonia is usually not fully conversed in the reactor. Higher inlet temperature of ammonia contributes to higher conversion rate. Therefore, when ammonia flows out of a liquid ammonia storage tank, ammonia needs to absorb substantial heat to ensure that it fully vaporizes and reaches a high temperature. The latent heat of evaporation is 23.3 kJ/mol at standard conditions. A simple estimation can be made: a total of 69.4 kJ of energy is required to vaporize and decompose 1 mol of ammonia. The hydrogen produced contains 360 kJ of energy (the low heat value of hydrogen is 120 kJ/g). However, the fuel energy cannot be fully converted into electric energy. Assuming a thermal efficiency of 40%, the hydrogen can produce 144 kJ of useful energy, signifying that at least 48.2% of the generated useful energy will be dedicated to electric heaters if the thermal requirement is met entirely by electric heaters, resulting in a significant reduction in the energy efficiency of the system.

To improve system efficiency, it is essential to recovery energy from residual heat sources. Ammonia decomposition reactors integrated with heat exchanger are mainstream choice for hydrogen production plants for ammonia-fueled internal combustion engines. One of the simplest designs is to place an ammonia decomposition reactor with the catalyst in the engine exhaust line [26]. However, there are several problems with this design. The first problem is the poor controllability of the exhaust gas from the internal combustion engine as a single heat source, which makes it difficult to obtain the desired decomposition rate. The second problem is the lack of hydrogen availability for cold starts. Reactors that use both electric heaters and combustion engine exhaust gases to provide heat can solve the cold start problem and improve the controllability of the system [27].

Energy efficiency analysis is important and necessary for automotive power systems. System efficiency reflects the practicability of automotive power systems. Determining the variation of system efficiency with the system output power helps in the development of energy management strategies, which are of great significance in improving the fuel economy of hybrid vehicles. The only attempt known to us to analysis ammonia-fueled automotive power systems efficiencies is conducted by Ezzat [21,23]. In Ezzat's paper, two integrated ammonia-fueled power systems, one uses an ICE and a PEMFC as engines and the other one without the PEMFC, are analyzed and compared thermodynamically utilizing energy and exergy.

However, previous studies have not focused on the impact of hydrogen production and energy recovery on ammonia-fueled power systems efficiencies. In this paper, three kinds of ammonia-fueled power systems are modeled and simulated to analyze their energy efficiencies. The hydrogen production process is considered in detail. The most significant factors impacting systems efficiencies are identified, and guidelines for system design and parameter optimization are proposed.

## 2. Systems configuration and description

This study focuses on three kinds of ammonia-fueled automotive configurations, which is shown in Fig. 1. The hybrid electric vehicle (HEV) consists of an ammonia-hydrogen-fueled ICE as the engine, a battery pack as the energy storage system, and an ammonia decomposition unit to produce hydrogen. The fuel cell electric vehicle (FCEV) consists of a PEMFC as the engine, a battery pack, an ammonia decomposition unit, and a hydrogen separation unit to purify hydrogen. The composite-powered electric vehicle (CEV) uses both an ammonia-hydrogen-fueled ICE and a PEMFC as the engines, along with a battery pack, an ammonia decomposition unit (ADU), and a hydrogen separation unit (HSU).

In this paper, we focus on the difference in output power and energy efficiency between engines and the system caused by hydrogen production and residual heat recovery. The power system is defined as a system comprising engines and components related to hydrogen production. The power system of the HEV is referred to in this paper as an ICE hybrid system, the power system of the FCEV is an FC hybrid system, the power system of the CEV is an composite power system.

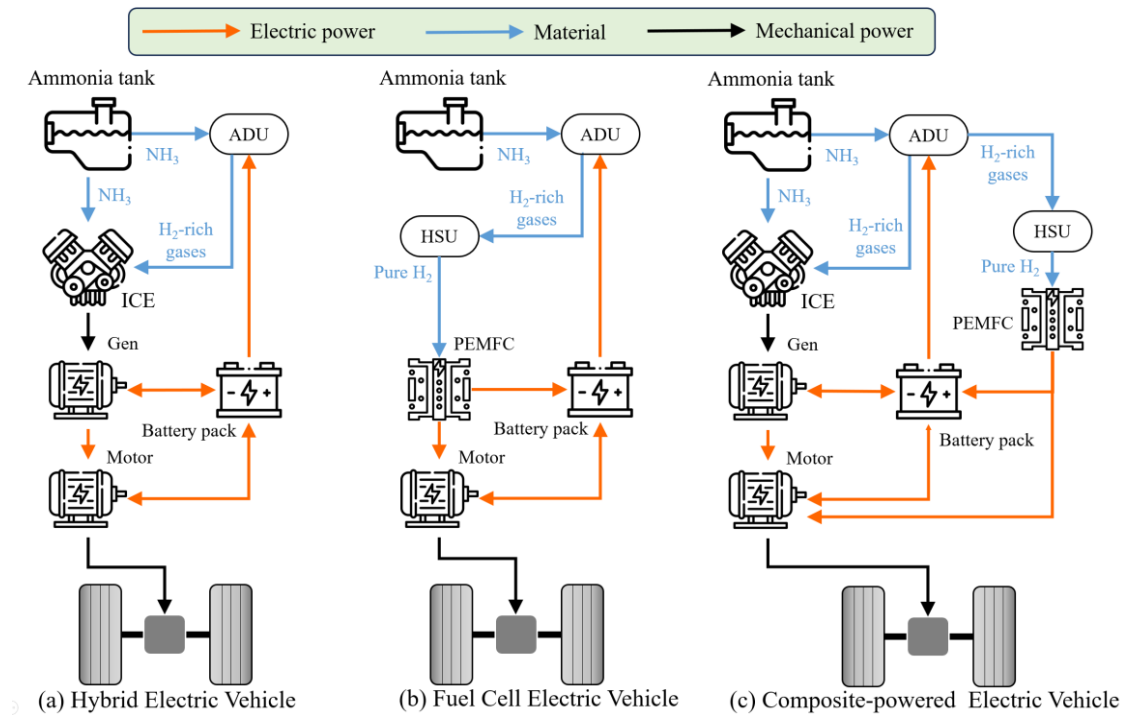

Fig. 1. Configurations of three kinds of ammonia-fueled power systems

## 2.1. Description of the composite power system

The composite power system is primarily described, as the ICE hybrid power system and the FC hybrid power system may be regarded as simplified versions of it.

- Power flow

Fig. 2 shows the power flow of the ammonia-fueled composite power system. The composite power system comprises three power sources: the ICE, the PEMFC and the battery pack. The ICE, which utilizes ammonia and hydrogen as fuel, converts the chemical energy in the fuels into mechanical energy. The generator transforms mechanical energy into electrical energy. The PEMFC converts the chemical energy in hydrogen into electrical energy directly through electrochemical reaction. Additionally, the system incorporates a battery pack for electrical energy storage. The battery pack serves as an energy reservoir, capable of storing electrical power when excess energy is available and releasing it as needed. This flexibility ensures a stable and reliable power supply for the system's various components and functions. Furthermore, it decouples the power output from engines from the vehicle's power demand, thereby reducing the engines' power fluctuations and making the engines as stable and efficient as possible. The electrical power is utilized to propel the vehicle and power auxiliary components, such as compressors for PEMFC, the ammonia pump for ICE, and electrical heaters.

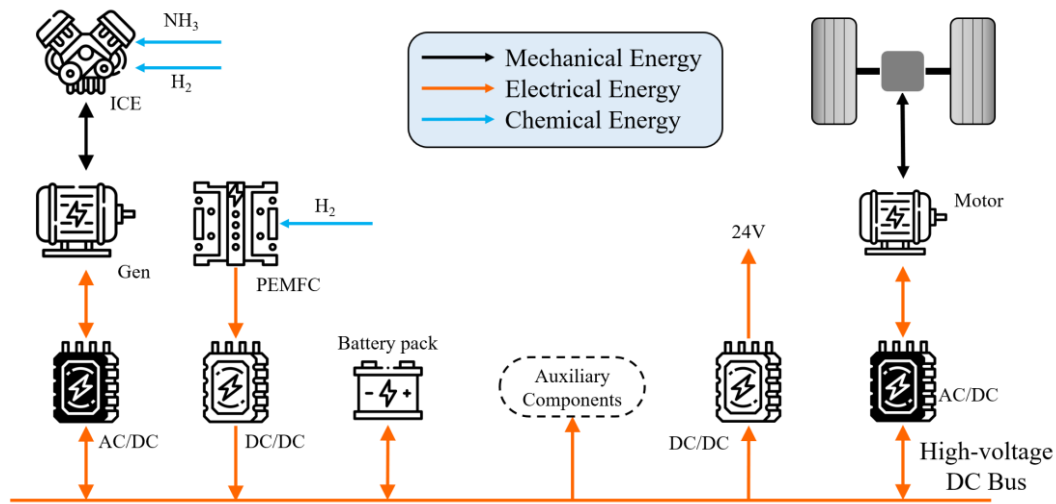

Fig. 2. Power flow of the ammonia-fueled composite power system

- Material flow

Fig. 3 shows the material flow of the ammonia-fueled composite power system. The primary fuel for the internal combustion engine is ammonia, with a portion of the ammonia from the tank delivered directly to the engine and the remainder entering the ammonia decomposition unit following preheating. In the decomposition unit, ammonia absorbs heat and decomposes into hydrogen and nitrogen, with the conversion rate dependent on the gas hourly space velocity and inlet temperature, which will be elucidated in greater detail subsequently. Subsequently, the hydrogen-rich gas, which has undergone decomposition, now contains hydrogen, nitrogen, and traces of ammonia. This gas then leaves the decomposition unit for further processing. The gas is divided into two separate paths. In one path, the hydrogen-rich gas is delivered to the internal combustion engine. In the second path, the hydrogen-rich gas is directed to the separation unit. The high-purity hydrogen obtained from the separation process is fed into the proton exchange membrane fuel cell, while the nitrogen-rich gas is fed into a reprocessing system.

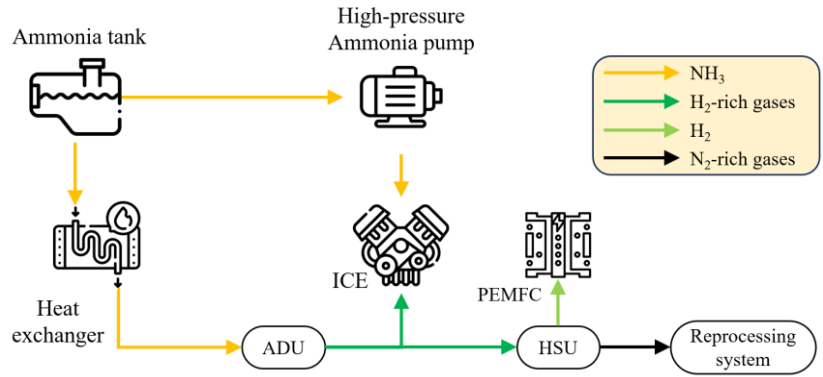

Fig. 3. Material flow of the ammonia-fueled composite power system

## 2.2. Qualitative analysis of system efficiency

The energy transformation and consumption of an ammonia-fueled power system is shown in Fig. 4. Engines convert the fuel's chemical energy into heat and useful output. In an ICE-Generator unit, fuel undergoes combustion with oxygen from the air, releasing chemical energy in the form of heat. The heat generated is converted into the rotational kinetic energy of the crankshaft ultimately, and a portion of the heat is carried away by exhaust gases and coolant, and lost due to friction and lubrication. The generator converts rotational kinetic energy into electrical energy, with some of the energy being dissipated as heat. In a PEMFC, the electrochemical reaction between hydrogen and oxygen produces electrical energy as the useful output and some heat as a loss. The efficiency of the engine for both the internal combustion engine-generator unit and the PEMFC is the ratio of the final electrical energy produced to the chemical energy of the fuel consumed.

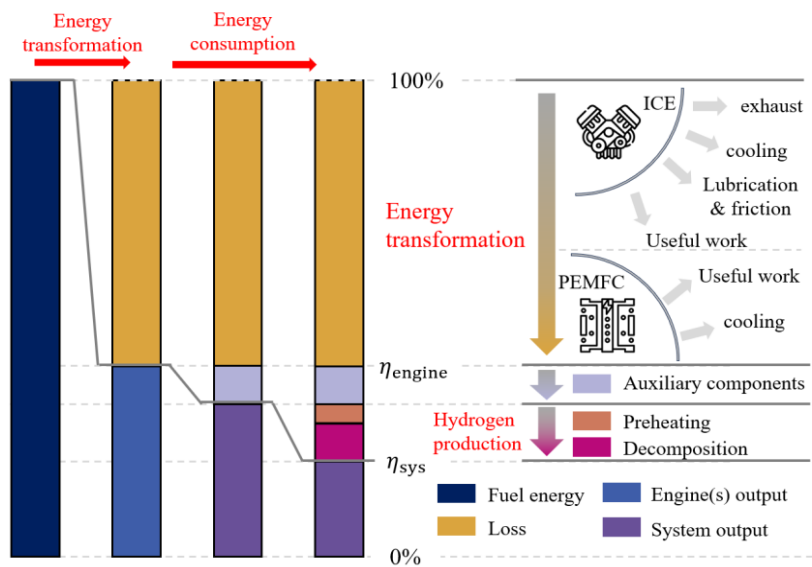

Fig. 4. Energy transformation and consumption of an ammonia-fueled power system

In power systems, electrical energy is required to operate essential auxiliary components like pumps and compressors to keep the engines running. This results in an unavoidable reduction in system energy efficiency. The reduced system efficiency is also affected by hydrogen production. Electric heaters consume electrical energy to provide heat for the preheating and decomposition process. The electrical energy generated by the engine is used for auxiliary components, and is consumed by preheating and decomposition.

In ammonia-fueled power systems, the system efficiency reduction caused by engine and hydrogen production is dependent on the type of engine. In general, FCs are more efficient at converting chemical energy into useful work, while ICEs are less-consuming at hydrogen production process. FCs are fed with high-purity hydrogen, results in a greater energy consumption during the preheating process and decomposition process.

**2.3. Approach to improve system energy efficiency**

Reducing energy consumption is clearly one way to improve the energy efficiency of the system. The energy consumption of hydrogen production is determined by the percentage of hydrogen in the fuel, the enthalpy change in the preheating process and the reaction enthalpy of the decomposition reaction. The percentage of hydrogen in the fuel is determined by engines. The enthalpy change in the preheating process is determined by decomposition temperature and pressure. The reaction enthalpy of the decomposition reaction, which is the chemistry of the reaction, is impossible to reduce. Therefore, using engines with a smaller percentage of hydrogen in the fuel and catalysts with higher conversion rates at lower temperatures is an effective way to reduce energy consumption, which in turn can improve system efficiency.

Recovering residual heat to reduce the energy consumption of electric heaters can also contributes to system efficiency. There are two key points in residual heat recovery: the amount of residual heat and the transfer of residual heat. The amount of residual heat is determined by engines, the decomposition unit and other components in the systems. The transfer of heat depends on the heat exchangers, the flow rate of the fluids and the temperature of the fluids. Therefore, the use of heat exchangers with higher heat transfer capacity, the design of a rational heat transfer sequence, and the improvement of the energy efficiency of the engine are ways to improve the energy efficiency of the system.

## 3. Residual heat recovery

Since both preheating and decomposition processes require heat, recovering the residual heat effectively reduces the consumption of the electrical heaters and thus improves the efficiency of the systems. In this section, the residual heat that can be utilized in systems are introduced and classified according to temperature first. Then, based on the characteristics of the decomposition reaction and the characteristics of the system residual heat sources, four different residual heat recovery measures are defined to compare the characteristics of the three systems with the differences in the subsequent analyses. In this study, only the amount of residual heat energy is considered, and exactly how to design the heat exchanger to recover the residual heat more fully will be further explored in future studies.

### 3.1. Residual heat sources in systems

The system contains various forms of residual heat that can be effectively utilized. Residual heat in the system can be divided into residual heat generated by the engines and residual heat from intermediate products.

The residual heat generated by the engine that can be recovered and utilized is the heat in the ICE exhaust gases, the heat in the engine coolant and the heat in the PEMFC coolant. Exhaust gases from the ICE have high temperatures, up to about 600°C[18]. The temperature, specific heat and composition of exhaust gases depend on the operating conditions and the specific design of the internal combustion engine.

Residual heat from intermediate products: some of the intermediate products do not require higher temperatures or even have to be cooled down for the next step of the reaction, so this part of the residual heat can also be utilized. The hydrogen-rich gas produced by the ADU has a high temperature, but it doesn't need to be at a high temperature whether it's going into the internal combustion engine or the HSU, so this residual heat can be utilized, and the same for the hydrogen and nitrogen-rich gas coming out of the HSU.

### 3.2. Definition of residual heat recovery measures

Residual heat that can provide heat for the decomposition process directly through the heat exchanger must be at a temperature higher than the decomposition temperature, so it can be categorized according to whether the temperature is higher than the decomposition temperature. Only the temperature of the exhaust gas from the internal combustion engine in the system may exceed the decomposition temperature, so only its heat can be used for the decomposition process. The remaining residual heat can only be utilized in the preheating process, in case direct heat exchange through a heat exchanger is considered.

In this study it was assumed that enough heat can be recovered to meet the demand, when the heat needed is not more than residual heat. In other words, only the relationship between the amount of residual heat and the amount of demanded heat is considered in this study, and exactly how to maximize or greater degree of residual heat recovery by designing the heat exchanger is not considered in this study.

Based on the different heat sources of the two processes, four measures of residual heat recovery are defined.

Measure I: None residual heat is recovered. The heat needed for the ammonia preheating and decomposition processes is solely provided by electric heaters. In this measure, the residual heat within the system remains untapped for these processes and is therefore considered the loss. The system energy efficiency in this case represents the lower limit.

Measure II: Low-temperature residual heat is recovered. In this measure, low-temperature residual heat is recovered and used for the preheating process. The heat required for the decomposition process is provided by electrical heaters. A portion of the engine output is needed to meet the energy requirements of the decomposition process.

Measure III: High-temperature residual heat is recovered. In this measure, when the heat required in the decomposition process is greater than that of the high temperature residual heat, the amount of residual heat utilized depends on the amount of high-temperature residual heat, and the electric heaters meets the remaining heat demand. All low temperature residual heat is not recovered.

Measure IV: Both low-temperature and high-temperature residual heat are recovered. High-temperature residual heat is used preferentially to meet the heat demand of the decomposition process; if it is more than the heat demand of the decomposition process, the remainder can be used to meet the heat demand of the preheating process. The amount of residual heat recovered depends on the amount of high-temperature residual heat and the heat demand in the preheating process. The system energy efficiency in this case represents the upper limit.

Fig. 5 shows the system energy consumption and recovery of residual heat under four residual heat recovery measures. The figure shows the amount of residual heat that is utilized and the processes in which it is used. It is worth noting that the energy of high-temperature residual heat includes only the heat that can be emitted by the exhaust gas from its original temperature down to the decomposition temperature.

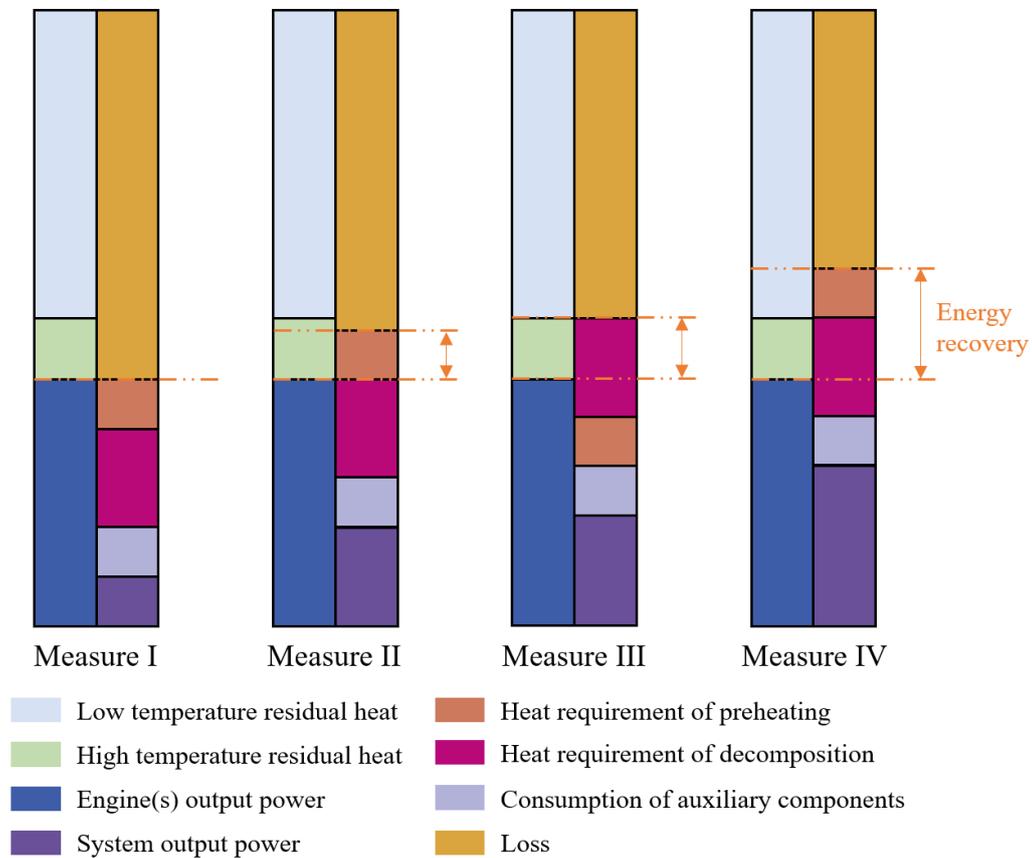

Fig. 5. system energy consumption and recovery of residual heat under four residual heat recovery measures

## 4. System energy efficiency modeling

In this section, some of the assumptions used in the modelling and simulation of the system are firstly clarified, and then the mathematical models of the key subsystems: the ammonia decomposition unit, the internal combustion engine and the proton exchanger membrane fuel cell are given separately. Then, the calculation of important evaluation metrics, including system energy efficiency and system output power, are explained.

### 4.1. Modeling assumptions

The following assumptions are made for system modelling and evaluation:

1. The reference temperature and reference pressure values are 298.15 K and 101.325 kPa, respectively.

2. The ammonia decomposition reactor is assumed to keep a constant temperature of 450°C.

3. The isentropic efficiency of the pump and compressor is 80%.

4. The hydrogen separation unit separates hydrogen completely and its energy consumption is neglected.

5. The combustion in the engine is considered complete.

### 4.2. Modeling of components

- Ammonia Decomposition Unit

The conversion ratio of ammonia can be expressed based on the dissociation fraction, $r$, so the decomposition reaction is expressed using following equation:

$$\mathrm{NH_3} \xrightarrow{\text{heat}} \frac{3r}{2}\mathrm{H_2} + \frac{r}{2}\mathrm{N_2} + (1-r)\mathrm{NH_3}$$

The dissociation fraction can be calculated using the following equation:

$$r = n_{\mathrm{NH_3}}^{\mathrm{out}} / n_{\mathrm{NH_3}}^{\mathrm{in}}$$

where $n_{\mathrm{NH_3}}^{\mathrm{out}}$ and $n_{\mathrm{NH_3}}^{\mathrm{in}}$ are the molar flow of inlet ammonia and outlet ammonia, respectively.

The reaction required heat can be expressed as follows:

$$\Delta h_{\mathrm{DU}} = \dot{m}_{\mathrm{NH_3,DUin}} \cdot r \cdot h_d$$

where $h_d$ is the heat essential for the endothermic reaction, kJ/kg.

The balance equation of ammonia decomposition unit can be expressed as followed:

$$\frac{dn_i}{dz} = v_i \cdot R \cdot A$$

where $n_i$ is the molar flow of the specie, $v_i$ represents the stoichiometric coefficient, $R$ is the reaction rate and $A$ is the cross-sectional area of the reactor.

The reaction rate of ammonia decomposition is expressed by the Temkine-Pyzhev model:

$$R = k_0 \cdot \exp\left(-\frac{E}{RT}\right) \cdot \left[\left(\frac{p_{NH_3}^2}{p_{H_2}^3}\right)^\beta - \frac{p_{N_2}}{K_{eq}^2}\left(\frac{p_{H_2}^3}{p_{NH_3}^2}\right)^{1-\beta}\right]$$

Where $k_0$ is the apparent pre-exponential factor, $E$ indicates the activation energy of the catalyst for NH$_3$ decomposition, $\beta$ represents the exponential constant related to the reaction order.

Table 1

Characteristics data of catalyst

| Parameter | Value |
| --- | --- |
| Activation energy of the catalyst, $E$ | 117 kJ/mol |
| Apparent pre-exponential factor, $k_0$ | 1.5e7 |
| Exponential constant, $\beta$ | 0.27 |

The values of the apparent pre-exponential factor $k_0$, the activation energy of the catalyst $E$ and the exponential constant $\beta$ are determined by catalyst. The characteristics data of catalyst used in this study is obtained from references [27,28] and shown in table 1.

Fig. 6 illustrates the variation of conversion rate and the hydrogen development rate with ammonia gas hourly space velocities when the inlet temperature of ammonia is maintained at 450°C. The data was obtained by simulation using Aspen software. The minimum catalyst volume necessary for the system is determined by the system's hydrogen requirements at maximum power output and the minimum acceptable conversion rate.

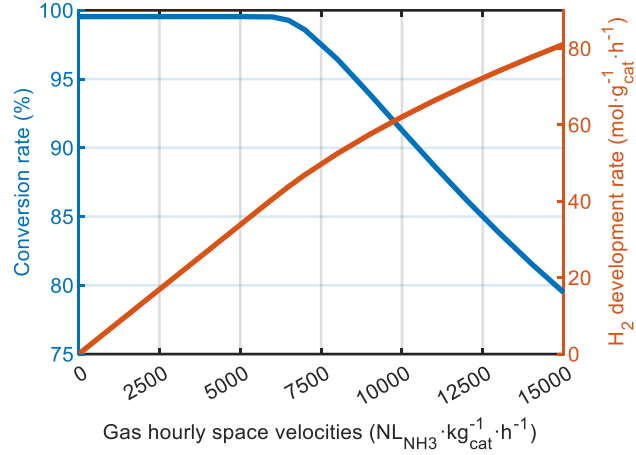

Fig. 6. Variation of the conversion rate and the hydrogen development rate with the ammonia gas hourly space velocities, 450°C.

- ICE-Gen Unit (IGU)

The parameters and operating conditions of the ammonia-hydrogen-ICE are based on Wang et al [29]. The data of the generator is from real tests.

Table 2
Parameters and operation conditions of the ammonia-hydrogen-fueled ICE

| Parameter | Value |
| --- | --- |
| hydrogen mole ratio, $a$ | 0.2 |
| Maximum thermal efficiency | 42.5% |
| Injection pressure of ammonia | 0.6 MPa |
| Injection pressure of hydrogen | 2.5 MPa |

When the hydrogen mole ratio is $a$ and the excess air ratio is $\lambda$, the combustion can be written as follows:

$$(1-a)\mathrm{NH_3} + a\mathrm{H_2} + \lambda(0.75 - 0.25a)(\mathrm{O_2} + 3.76\mathrm{N_2})$$
$$\rightarrow (1.5 - a)\mathrm{H_2O} + (3.76\lambda - 1.44a - 0.44)\mathrm{N_2} + (\lambda - 1)\mathrm{O_2}$$

The energy balance equation for the ICE is expressed using following equation:

$$\dot{m}_{\mathrm{NH_3,ICE}} h_{\mathrm{NH_3,ICE}} + \dot{m}_{\mathrm{H_2,ICE}} h_{\mathrm{H_2,ICE}} + \dot{m}_{\mathrm{air,ICE}} h_{\mathrm{air,ICE}}$$
$$= \dot{m}_{\mathrm{exh,ICE}} h_{\mathrm{exh,ICE}} + \dot{W}_{\mathrm{ICE}} + \dot{Q}_{\mathrm{cooling,ICE}} + \dot{Q}_{\mathrm{lub,ICE}}$$

where $\dot{W}_{\mathrm{ICE}}$ refers to the ICE obtainable power, $\dot{Q}_{\mathrm{cooling,ICE}}$ refers to the amount of heat rejected to the cooling system, $\dot{Q}_{\mathrm{lub,ICE}}$ refers to the heat loss due to friction and ICE lubrication system.

The output power of the generator:

$$\dot{W}_{FC} = \dot{W}_{ICE} \cdot \eta_{Gen}$$

The ICE-Gen unit always works on the optimal efficiency curve, which is shown in Fig. 7.

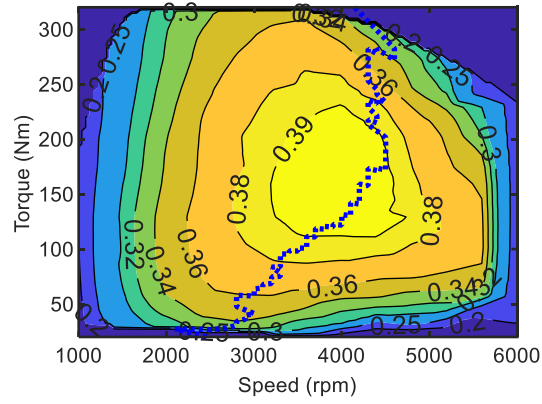

Fig. 7. Efficiency map of the ICE-Gen unit

- **Fuel cell**

The data of the proton exchange membrane fuel cell is from an example of MATLAB/Simulink.

The output voltage of the fuel cell can be obtained as follows:

$$V_{stack} = N_{cell} \cdot V_{cell}$$

where $N_{cell}$ represents cells number in the fuel cell stack, $V_{cell}$ refers to the practical cell potential.

The practical cell potential is expressed using following equation:

$$V_{cell} = E_N - \eta_{act} - \eta_{conc} - \eta_{ohm}$$

where $E_N$ represents the reversible cell potential, V; $\eta_{act}$ refers to the losses due to the activation overpotential, V; $\eta_{conc}$ represents the losses due to concentration overpotential, V; $\eta_{ohm}$ denotes the losses due to ohmic overpotential, V.

The power generated by the whole fuel cell stack is determined by the following equation:

$$\dot{W}_{FC} = N_{cell} \cdot V_{cell} \cdot i \cdot A_{cell}$$

where $i$ refers to the current density, $A/cm^2$; $A_{cell}$ represents the fuel cell area,

cm².

The reversible cell potential can be obtained as follows:

$$E_\text{N} = E_\text{T}^0 + \frac{RT}{2F}\ln\left(\frac{p_{H_2}}{p_0}\sqrt{\frac{p_{O_2}}{p_0}}\right)$$

where $E_\text{T}^0$ is the Nernst voltage at standard pressure, V; $R$ is the universal gas constant, 8.314 J/(mol · K); $T$ represent the temperature at which the cell operates, K; $F$ is the Faraday constant, 96485 C/mol; $p_0$ is the standard atmospheric pressure, 101.325 kPa; $p_{H_2}$ and $p_{O_2}$ are the partial pressure of hydrogen and oxygen that involved in the reaction, kPa.

The losses due to the activation overpotential characterizes the activation energy barrier that needs to be overcome to carry out an electrochemical reaction and is related to the current density. It may be determined as follows:

$$\eta_\text{act} = \frac{RT}{2\alpha F}\ln\left(\frac{i}{i_0}\right)$$

where $\alpha$ is the charge transfer coefficient, $i_0$ refers to the exchange current density, A/cm².

The losses due to concentration overpotential characterizes the voltage loss due to the concentration of reactants on the catalyst surface being lower than the total concentration of reactants. It can be obtained as follows:

$$\eta_\text{conc} = \frac{RT}{2F}\ln\left(\frac{i_L}{i_L - i}\right)$$

where $i_L$ represents the max (limiting) current density, A/cm².

The losses due to ohmic overpotential is calculated as follows:

$$\eta_\text{ohm} = iR_\text{ohm}$$

$$R_\text{ohm} = \frac{L_\text{m}}{\sigma_\text{m}}$$

where $L_\text{m}$ is the membrane thickness, cm; $\sigma_\text{m}$ refers to the conductivity of the membrane, $\Omega^{-1} \cdot \text{cm}^{-1}$.

### 4.3. Evaluation metrics

- System output power

The system output power is the total power generated by the system minus the power of the components that are necessary to keep the system in proper working order.

The system output power is calculated as follows:

$$\dot{W}_{sys} = \dot{W}_{Gen} + \dot{W}_{FC} - \dot{W}_{pump} - \dot{W}_{comp,air} - \dot{W}_{EH}$$

where $\dot{W}_{Gen}$ and $\dot{W}_{FC}$ are the power generated by the generator and the PEMFC, $\dot{W}_{pump}$, $\dot{W}_{comp,air}$ and $\dot{W}_{EH}$ are the power consumed by the ammonia pump, air compressor and electric heaters, respectively.

- Energy efficiency

The energy efficiency of the ICE-Gen unit is written as shown in the following equation:

$$\eta_{ICE-Gen} = \frac{\dot{W}_{Gen}}{\dot{m}_{NH_3,ICE}LHV_{NH_3} + \dot{m}_{H_2,ICE}LHV_{H_2}}$$

where $LHV_{NH_3}$ and $LHV_{H_2}$ are the low heat values of ammonia and hydrogen.

The energy efficiency of the FC is written as shown in the following equation:

$$\eta_{FC} = \frac{\dot{W}_{FC}}{\dot{m}_{H_2,FC}LHV_{H_2}}$$

The energy efficiency of the system is written as follows:

$$\eta_{sys} = \frac{\dot{W}_{sys}}{\dot{m}_{NH_3}LHV_{NH_3}}$$

## 5. Results and discussion

In this section, the simulation results of the above three systems are shown and compared. The discussion starts with the comparison of two important evaluation metrics: energy efficiency and system output power under four measures of residual heat recovery. Then, by analyzing the energy transformations of the three systems, it is shown how the differences in the engines between the systems lead to differences in energy efficiency and maximum. After that, the simulation results of the three systems under driving cyclic test conditions are used to analyze how the energy efficiency and the maximum system output power affect the driving of the vehicle. Finally, it is explored how to determine the maximum output power of the ICE and the PEMFC in the composite power system in order to obtain higher energy efficiency provided that the system output power is sufficient.

## 5.1. Simulation results of three kinds of systems

### 5.1.1. Ammonia-fueled ICE hybrid system

Fig. 8 shows the variation of the maximum energy efficiency and system output power with the output power of ICE. The ICE-generator co-generation unit achieved a maximum efficiency of 39.34% at ICE output of 89.5 kW. The ammonia pump led to a slight drop, resulting in a maximum efficiency of 38.78%. In measure III, where the thermal requirement of preheating process is met by electric heaters, the maximum efficiency decreased to 36.48%, indicating a 2.30% reduction in system efficiency. Similarly, measure II, where the thermal requirement of decomposition process is met by electric heaters, showed a maximum efficiency of 37.89%, resulting in a 0.89% decline in system efficiency. Finally, measure I, encompassing both processes of hydrogen production, showed a maximum efficiency of 35.59%, reflecting a total decrease in system efficiency of 3.19%. With full recovery of both low-temperature and high-temperature residual heat, the system achieves a maximum energy efficiency of 38.78% with a corresponding system output power of 89.5 kW. The system's maximum system output power reaches 215.8 kW, resulting in an energy efficiency of 28.83%. In contrast, when residual heat is not utilized at all, the maximum energy efficiency is 35.59%, yielding a system output power of 91.00 kW. The system's maximum system output power in this scenario is 191.2 kW, corresponding to an energy efficiency of 25.55%.

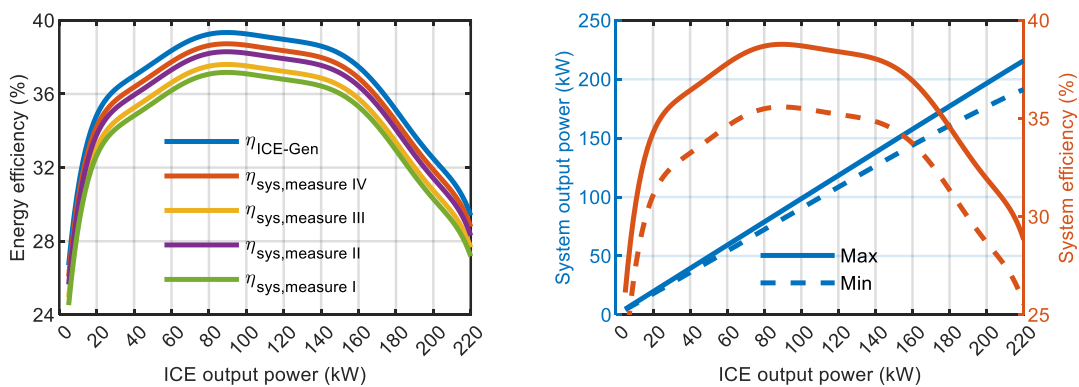

(a) (b)

Fig. 8. Variation of energy efficiency (a) and system output power (b) with the output power of the ICE

### 5.1.2. Ammonia-fueled FC hybrid system

Fig. 9 shows the variation of the energy efficiency and system output with the output power of FC. In this system, the fuel cell achieves a maximum energy efficiency of 65.00% when its output is at a minimum of 1 kW. However, the inclusion of the air compressor in the separation unit reduces the maximum energy efficiency to 59.87%. Since there is no residual heat source above 450°C for the decomposition process, the system attains a maximum energy efficiency of 45.29% when residual heat sources provide heat for the preheating process. When electric heaters meet the preheating process's heat demand, the system's maximum energy efficiency decreases to 39.68%. Hydrogen production resulted in a 20.19% decrease in the maximum energy efficiency of the system, while the recovery of residual heat can improve it by up to 8.61%. With full recovery of residual heat, the system achieves a maximum energy efficiency of 45.29% with a corresponding system output power of 10.52 kW. The system's maximum system output power reaches 112.28 kW, resulting in an energy efficiency of 21.72%. In contrast, when residual heat is not utilized at all, the maximum energy efficiency is 39.68%, yielding a system output power of 9.82 kW. The system's maximum output power in this scenario is 83.90 kW, corresponding to an energy efficiency of 19.15%. At higher output power, more energy is used for hydrogen production due to a decrease in energy efficiency of fuel cell and a decrease in the conversion rate of decomposition unit, leading to a decrease in system output power.

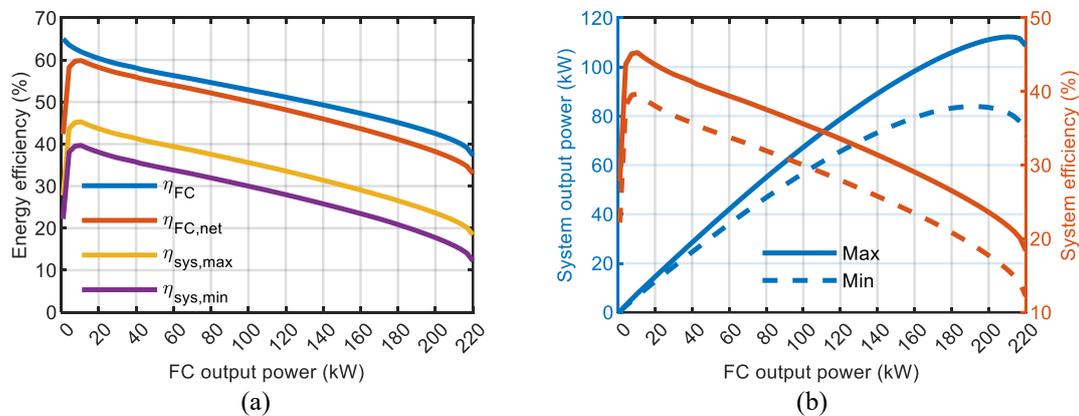

(a) (b)

Fig. 9. Variation of energy efficiency (a) and system output power (b) with the output power of the fuel cell

### 5.1.3. Ammonia-fueled composite power system

Fig. 10 shows the efficiency map of system in measure I (a), measure II (b), measure III (c) and measure IV (d), the blue dashed curves in the figure are the iso-power curves of the system, and the red dash-dotted curves are the optimal power distribution curves. From measure I to measure IV, as the residual heat recovery rate increases, the high efficiency zone and the maximum energy efficiency of the system become larger. The iso-power curves become denser indicating that more system output power. Residual heat recovery rate also has a large impact on the power distribution of the system, as residual heat recovery rate increases, the greater the proportion of FC output to total output power, the later the ICE goes to full load. When system power requirements are low, using only fuel cell is the more efficient option. In measure I, without utilizing residual heat, it is efficient to make ICE output more power than FC. In measure II, where the heat requirement of preheating process is met by residual heat sources, the optimal power distribution curve is higher than that in measure I. This means that for the same power requirement, having more output from the fuel cell makes the system more efficient. In measure III and measure IV, where the exhaust gases can be used to provide heat for decomposition process, the ratio of power output from ICE and FC tends to a stable value, as more high-temperature residual heat from a larger ICE output allows the decomposition unit to produce more hydrogen for the fuel cell.

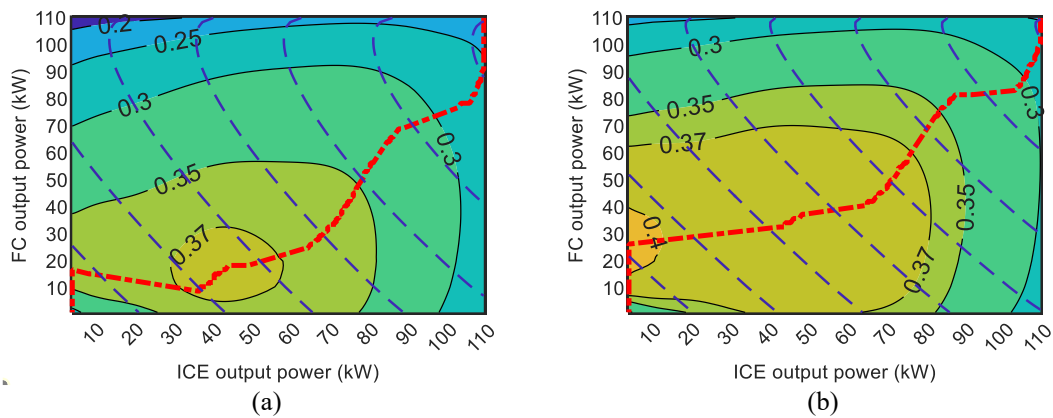

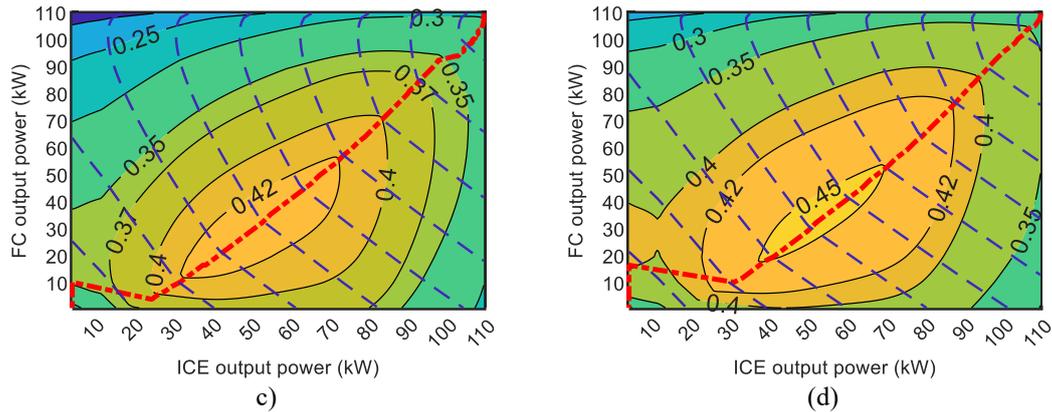

Fig. 10. Efficiency map of the composite power system in measure I (a), measure II (b), measure III (c) and measure IV (d)

Fig. 11 shows the optimal efficiency curves in different measures. These curves show the highest energy efficiency of the system at each system output power. As can be observed from the figure, the hydrogen production and the recovery of residual heat have a significant impact on the energy efficiency of the system, also have an impact on system output power. The extensive recovery of residual heat contributes to increase the energy efficiency of the system, resulting in expanding the high-efficiency zone of the system and greater maximum system output power. The maximum energy efficiency of system in four measures are 39.91%, 40.71%, 43.32%, and 45.72%, when the system output power is 44.0 kW, 22.0 kW, 69.9 kW, and 79.6 kW, respectively. The maximum system output power in four measures is 141.9 kW, 160.9 kW, 189.3 kW, and 210.3 kW, with the system output power is 24.18%, 25.88%, 29.85%, and 32.03%, respectively. The increased recovery rate of residual heat enhances the maximum energy efficiency of the composite power system by 5.81% and maximum system output power by 68.4 kW.

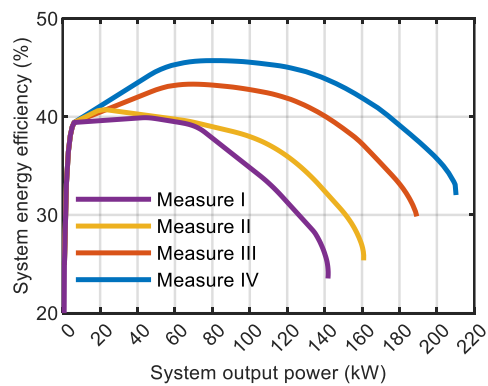

Fig.11 Optimal efficiency curves in different measures

## 5.2. Comparison of the simulation results of three kinds of systems

### 5.2.1. Comparison of energy efficiency and maximum system output power

Fig. 12 compares the three systems in terms of system energy efficiency and maximum system output power under different measures of residual heat recovery. The benefits of increased residual heat recovery rate are more pronounced for the composite power system, where both maximum energy efficiency and maximum system output power increase much faster compared to the ICE hybrid system, also resulting in higher energy efficiency over a wider power range.

Maximum system output power: the maximum system output power of the ICE hybrid system is greater than that of the composite power system is greater than that of the FC hybrid system. The results indicate that, when the maximum output power of engines is equivalent, the reduction in fuel cell output from hydrogen production cannot be compensated for by residual heat recovery, even under measure IV where the residual heat is fully utilized. In order for the outputs of three systems to be equivalent, the FC hybrid system requires a more powerful engine, while the ICE hybrid system requires an engine with minimal output power.

Maximum energy efficiency: the figure illustrates that the maximum energy efficiency of the ICE hybrid system is the lowest in all four measures. In measure I where the residual heat is not utilized at all and measure II where residual heat is used for the preheating process, the maximum energy efficiency of the FC hybrid is higher than that of the composite power system. However, in measure III and measure IV, the maximum energy efficiency of the composite power system is higher than that of the FC power system. This implies that the high efficiency advantage of the fuel cell renders the maximum energy efficiency of the FC hybrid system the highest, despite the fact that hydrogen production has a significant impact on the energy efficiency of the system.

Variation of the energy efficiency with the system output power: the FC hybrid system is more energy efficient at lower system. As the system output power increases, the composite power system demonstrates a clear advantage in terms of energy efficiency.

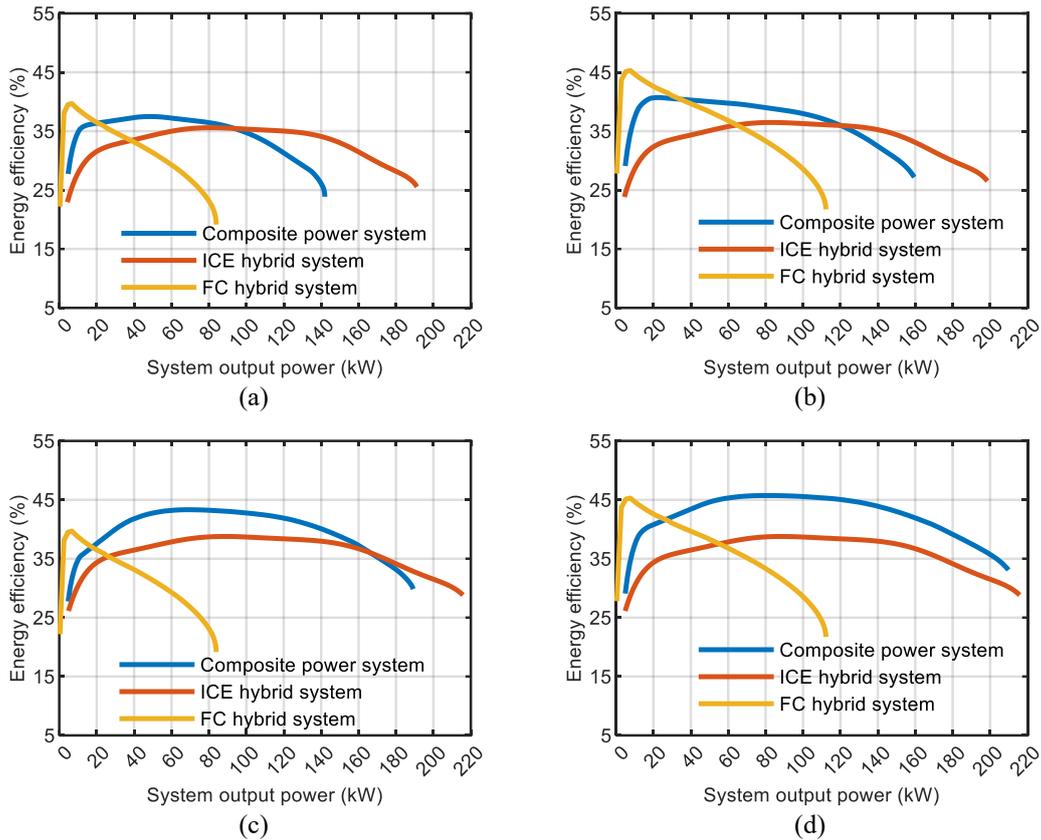

Fig.12 Variation of system energy efficiency of three systems with system output power in measure I (a), II (b), III (c) and IV (d)

### 5.2.2. Analysis of energy transformation and consumption

Fig. 13 shows the energy transformation and consumption of three systems at maximum system energy efficiency point and maximum system output power point in measure IV. As can be seen in Fig. 13, the difference in system energy efficiency is caused by a combination of three factors: engine energy efficiency, residual heat, and hydrogen production energy consumption.

Engine energy efficiency: the PEMFC has the highest energy efficiency, 62.3% and 40.7% respectively, at two operating points. The ICE, on the other hand, has the lowest energy efficiency at 40.0% and 30.0% at two operating points, respectively. That of the composite power system between them, 46.9%% and 33.5%, due to the combined use of the ICE and the PEMFC. Residual heat: the amount of residual heat is related to the energy efficiency of engines, and temperature is related to the energy transformation. There is no high-temperature residual heat source in the FC hybrid system, so the residual heat, 37.7% and 59.3%% of energy, is low-temperature residual heat in two points, respectively. In the ICE hybrid system, the exhaust gases from the ICE are high-temperature residual heat sources, with 8.0% and 13.8% of energy from high-temperature residual heat sources and 52.0% and 56.2% of energy from low-temperature residual heat sources, respectively. In the composite power system, the energy share of the exhaust gases from the ICE is 6.3% and 7.7%, and the energy share of the low-temperature residual heat from both the ICE and PEMFC is 46.8% and 58.8%, respectively. Hydrogen production energy consumption: the FC hybrid system has the most energy consumption of hydrogen production, totally 20.2% and 20.6%, respectively. In the ICE hybrid system, these values are 3.1% and 3.3%. The energy consumption of the composite power system is between other two systems, at 8.7% and 11.2%, respectively.

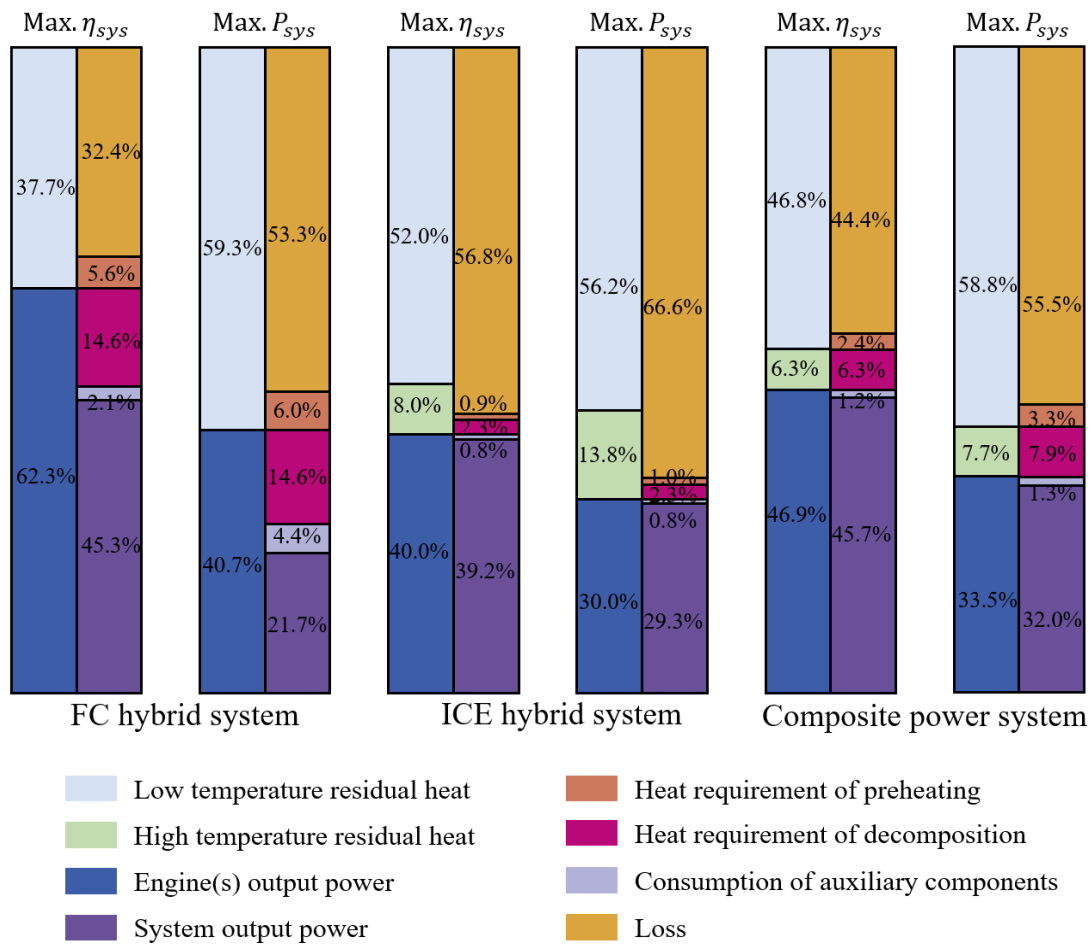

Fig.13 Energy transformation and consumption of three systems at maximum energy efficiency and maximum system output power

Despite the FC hybrid system has the highest engine energy efficiency, its system energy efficiency is the lowest of the three systems. This is due to the absence of high-temperature residual heat in the system that can be utilized for decomposition, and the high energy consumption required for hydrogen production. There is no high-temperature residual heat source because of the energy conversion method, direct electrochemistry. Therefore, the heat required for the decomposition process must be met exclusively by consuming electrical energy. The high energy consumption of hydrogen production is because the only fuel for PEMFC is hydrogen. The quantity of ammonia that must be decomposed is considerable, thus hydrogen production consumes substantial energy.

The energy efficiency of the ICE hybrid system is higher and acceptable. Even though the engine energy efficiency of the ICE hybrid system is lower than that of the FC hybrid system, the requirement of hydrogen is less and the high-temperature residual heat is more. Combustion in the ICE produces high-temperature exhaust gases, which can be used as a high-temperature residual heat source, to provide heat for the decomposition process. The main fuel of the ammonia-hydrogen-fueled ICE is ammonia, and hydrogen accounts for about 2.86% of the fuel. The heat required for decomposition is less, the residual heat can theoretically meet the heat demand, so that hydrogen production does not lead to a decrease in efficiency, and the energy efficiency of the system is basically equal to that of the ICE. However, the residual heat in the exhaust gases of the ICE is not fully utilized.

For the composite power system, it combines the advantages of the PEMFC and the ICE. By rationally distributing the output power of the ICE and the PEMFC, the energy from the high-temperature residual heat source in the system can be fully utilized for hydrogen production. The PEMFC can consume more hydrogen, thereby enabling the generation of more power and an increase in the total engines energy efficiency. Furthermore, it can achieve a higher system energy efficiency.

### 5.3. Parameter schemes of composite power system

The maximum output power of the ICE and PEMFC affect the system energy efficiency and the system maximum output power in a composite power system. To compare the differences between systems with different parameter schemes, the measure where the residual heat is fully utilized (measure IV) is set. Define the ratio of maximum output power of the ICE to the total maximum output power of engines as $r_{ICE}$.

$$r_{\text{ICE}} = \frac{P_{\max}^{\text{ICE}}}{P_{\max}^{\text{ICE}} + P_{\max}^{\text{FC}}}$$

Fig. 14 shows the variation of system maximum energy efficiency and maximum system output power with the ratio of the maximum output of the ICE to the total maximum output of engines. As $r_{\text{ICE}}$ increases from 0.1 to 0.9, the maximum energy efficiency of the system decreases from 47.21% to 40.64% and the maximum system output power increases from 115.4 kW to 208.6 kW.

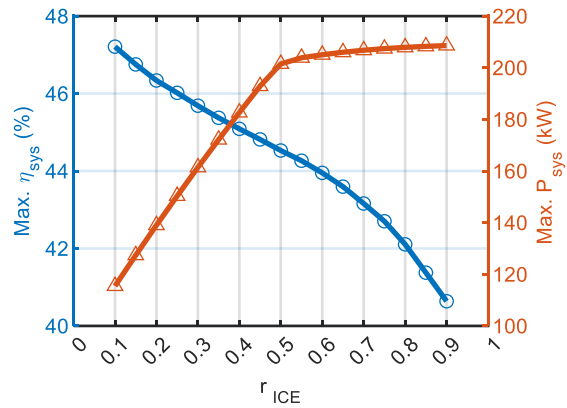

Fig. 14. Variation of system maximum energy efficiency and maximum system output power with the $r_{\text{ICE}}$

Fig. 15 shows the variation of system energy efficiency with the output power of the system. It can be seen from this figure that the system output power corresponding to the maximum energy efficiency is increasing as $r_{\text{ICE}}$ increases, which means that the load factor to reach the maximum energy efficiency is higher as $r_{\text{ICE}}$ increases. The system output power corresponding to the maximum energy efficiency is 62.0 kW, 86.9 kW and 88.8 kW, when $r_{\text{ICE}}$ is 0.25, 0.5 and 0.75, respectively. Corresponding load factors are 37.33%, 41.20% and 41.74% respectively.

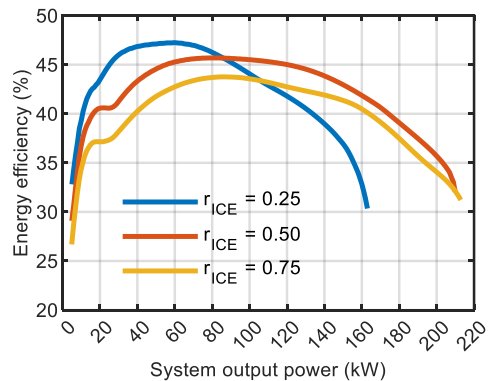

Fig. 15. Variation of system energy efficiency with system output power

As $r_{ICE}$ increases, the maximum output power of the PEMFC decreases, and the maximum value of hydrogen required by the system, which mainly affected by PEMFC, decreases, so less energy is consumed for hydrogen production and the system maximum output power increases. Then, the maximum system energy efficiency. By analyzing the energy transformation and consumption at different $r_{ICE}$, it can be known that the main reason for the decrease in the maximum energy efficiency is the decrease in the efficiency of engines. Combined with the above analysis, it can be known that the optimal efficiency of the system must be reached when the energy of exhaust gas from the ICE is just enough to satisfy the decomposition heat demand. As $r_{ICE}$ decreases, the maximum energy that can be provided by the exhaust gas of the ICE decreases, and the less hydrogen can be produced by utilizing this energy, corresponding to the smaller output power of the PEMFC. According to the relationship between the energy efficiency and the output power of the PEMFC, the smaller the output power, the higher the energy efficiency. Therefore, the smaller the $r_{ICE}$, the higher the energy efficiency of the system engines.

Combining the contents of the two figures above, the principle of matching the power system parameters of a vehicle that uses this composite powertrain can be given: first, determine the desired system output power corresponding to the maximum energy efficiency, which is typically the power demand when the vehicle is cruising. Then, the maximum output power of the power system is determined, and it includes the backup power. Based on the maximum output power a series of parameter schemes can be obtained. Finally, based on the ratio of these two powers, i.e., the load factor at the point of maximum energy efficiency, the ratio of the maximum output power of the internal combustion engine to the maximum output power of engines, i.e. $r_{ICE}$ can be selected, which in turn determines the maximum output power of the ICE and the PEMFC.

## 6. Conclusion

The use of ammonia fuel in automotive power systems rely on an ammonia-hydrogen-fueled internal combustion engine or a proton exchange membrane fuel cell. These two types of engines in question convert the chemical energy in the ammonia fuel into mechanical or electrical energy through combustion or electrochemistry. Both ICEs and FCs require hydrogen, and on-board hydrogen production from ammonia decomposition is an energy-consuming process that reduces system efficiency. The ICE utilizes a small amount of hydrogen as the combustion enhancer, while the PEMFC require pure hydrogen as fuel. Consequently, the selection of engines in a power system configuration determines the extent to which hydrogen production affects the efficiency of the system.

For the ammonia-fueled ICE hybrid power system, the proportion of hydrogen in the fuel is relatively low, which results in a reduction in the energy required for hydrogen production. Furthermore, the energy efficiency reduction caused by hydrogen production is relatively minor, with a maximum of 3.19%. In the case of the FC hybrid power system, where hydrogen is the only fuel, the energy required for hydrogen production is higher and has a greater impact on the energy efficiency of the system, at most 23.19%.

The energy efficiency of a system can be improved through energy recovery by utilizing residual heat, the extent of which depends on the residual heat in the system and the degree of recovery. Since there are high-temperature exhaust gases in the ICE hybrid power system that can provide heat for the ammonia decomposition process, residual heat recovery can eliminate the impact of hydrogen production on energy efficiency. While the FC hybrid power system does not contain a high-temperature residual heat source, the heat required for the decomposition process can only be provided by electric heaters, and residual heat recovery improves energy efficiency by up to 8.61%, while hydrogen production still leads to a 14.58% decrease in energy efficiency. Therefore, using a PEMFC as the sole engine of an ammonia-fueled power system is not feasible from an energy efficiency point of view, while using only an ICE is feasible.

The composite power system using both an ICE and a PEMFC as engines has the highest energy efficiency of the three systems. Through power distribution, the residual heat energy in the exhaust gas of the ICE can be fully utilized to produce hydrogen, thus allowing the fuel cell to take full advantage of its high energy efficiency.

The ratio of the maximum output power of the ICE to the maximum output power of engines affects the maximum energy efficiency of the system and its corresponding system output power, the maximum system output power, and the high efficiency region of the system different. Based on the influence, the optimal point of maximum energy efficiency and maximum output power can be achieved for a composite power system by matching the parameters.

**Acknowledgment**

We would like to acknowledge the funding support of the National Natural Science Foundation of China (Grant No.52272372 and Grant No.52302456 ).